 \documentstyle[prl,aps]{revtex}   %%Journal Style
  \def\ack{\,|\,}
\begin{document}
\draft

%% Include following two lines for Journal Style

 \twocolumn[\hsize\textwidth\columnwidth\hsize  %**Journal
 \csname @twocolumnfalse\endcsname              %**Journal

\title{
Band Crossing and Signature Splitting in Odd Mass $fp$ Shell Nuclei
}

\author{Victor Vel\'azquez$^1$, Jorge G. Hirsch$^2$ and Yang
Sun$^{3,4}$}

\address{
$^1$ Groupe de Physique Th\'eorique, Centre de Recherches
Nucl\'eaires, \\
(IN2P3-CNRS-Universit\'e Louis Pasteur) B\^atiment 40/I, 
F-67037 Strasbourg Cedex 2, France\\
$^2$ Instituto de Ciencias Nucleares, Universidad Nacional
Aut\'onoma de M\'exico,\\ 
Circuito Exterior C.U., A.P. 70-543, 04510 M\'exico D.F., M\'exico\\
$^3$ Department of Physics and Astronomy, University of Tennessee,
Knoxville, TN 37996, U.S.A.\\
$^4$ Department of Physics, Xuzhou Normal University,
Xuzhou, Jiangsu 221009, P.R. China\\ 
}
\maketitle

\begin{abstract}
Structure of two sets of mirror nuclei: $^{47}$V - $^{47}$Cr and
$^{49}$Cr - $^{49}$Mn, as well as $^{49}$V and $^{51}$Mn, is studied using the
projected shell model. 
Their yrast spectra are described as an interplay between the angular
momentum projected states around the Fermi level which  
carry different intrinsic $K$-quantum numbers. 
The deviations from a regular rotational sequence 
are attributed to 
band crossing and signature splitting,
which are usually discussed in heavy nuclear systems. 
Our results agree reasonably with 
experimental data, and are comparable with those from the full $pf$ shell
model calculations. 
\end{abstract}


%%  Include the following line for Journal style
  ]  %**Journal

%\narrowtext
 \newpage

\section{Introduction}

Our knowledge about the nuclear structure around $^{48}$Cr has 
been improved enormously in recent years. Multi-element gamma-ray detectors
have allowed the establishment of rotational structures 
in $^{46,47}$Ti, $^{47,48}$V,
and $^{47,48,49,50}$Cr, with clear
evidence of backbending found in $^{48,50}$Cr
\cite{Cam94,Cam96,Cam98,Lenzi97,Bra98}. These experiments have been performed in
a close interaction with the full {\em pf} shell model calculations, which
have provided us with 
a nearly perfect description of the rotational bands in many of
these nuclei \cite{Cau94,Cau95,Pin96,Pin97}. 

While the backbending mechanism in heavy nuclei is commonly 
understood as a band
crossing phenomenon 
involving two rotational bands with different moments of inertia
\cite{Rin80}, the origin of backbending in $^{48,50}$Cr at
spin $I\approx 10$ has been debated. Caurier {\it et al.} 
\cite{Cau95} provided exact solutions of the Hamiltonian within the 
complete {\em pf} shell. The backbending effect was attributed to
a collective to non-collective transition. 
Later, Mart\'\i nez-Pinedo {\it et al.} 
\cite{Pin96} performed a similar study for $^{50}$Cr, in which they
reproduced the first backbending, 
and predicted a second backbending at $I=18$ which was confirmed 
experimentally one
year later \cite{Lenzi97}. Using a self-consistent cranked
Hartree-Fock-Bogoliubov formalism, Tanaka {\it et al.} offered 
an explanation for the backbending in $^{48}$Cr, in  
which they proposed that the phenomenon  
is not associated with a single-particle level crossing
\cite{Tan98}. 

In a recent paper by Hara {\it et al.} \cite{Hara99}, the backbending 
mechanism of $^{48}$Cr was studied within the projected shell model (PSM)
\cite{Hara95}, which has been successful in describing 
well deformed heavy nuclei \cite{Hara95} and those of  
transitional region \cite{Shei99}. 
In \cite{Hara99}, it was concluded that the backbending in
$^{48}$Cr is due to a band crossing involving an excited band 
built on simultaneously broken pairs of
neutrons and 
protons in the
``intruder''
subshell $f_{7/2}$. This type of band crossing is 
usually known to cause
a second
backbending in rare-earth nuclei.
In the same work \cite{Hara99}, 
it was shown, by using the generator coordinate method (GCM) 
with the same Hamiltonian as that of the $pf$ shell model
calculation,
that the backbending in $^{48}$Cr can be interpreted as due to
the crossing between the deformed and spherical bands.
Detailed analysis indicated that the two theories 
(PSM and GCM) lead to a consistent
picture of band crossing since the physical content of the 4-qp band
in PSM and that of the spherical band in GCM is very similar \cite{Hara99}.  

The purpose of the present paper is to demonstrate that the PSM is not only  
able to explain the backbending phenomenon in the even-even 
nucleus $^{48}$Cr, but it can, using the same 
set of parameters, describe the odd-mass nuclei $^{47,49}$V,$^{47,49}$Cr
and $^{49,51}$Mn. 
These nuclei have been extensively studied with the full $pf$ shell
diagonalizations \cite{Pin97}. 
It will be seen here that the band crossing mechanism
also explains the deviation 
from a regular sequence for rotational bands 
in some of these odd-mass nuclei. 
Furthermore, the observed energy staggering along a band can be well described
by the PSM in terms of the signature splitting, 
a terminology that originates from the particle-rotor model. 

The paper is arranged as follows: 
In Section II, we outline the PSM and 
present the Hamiltonian and the model space used in our calculation.
In Section III, we introduce the concept of signature splitting  
and show a general rule of energy staggering in a rotational band 
having quantum numbers $j$ and $K$. 
Detail discussion on the structure of these nuclei 
and comparison with experimental
data are given in Section IV.  
Finally, we summarize our paper in Section V.

\section{Outline of the Theory}

The use of a deformed basis allows construction of
an optimal set of basis states, in which shell model 
truncation can be done most efficiently
by selecting the low-lying configurations around the Fermi level 
\cite{Hara95}. 
It provides 
a good classification scheme in the sense that a simple configuration
corresponds (approximately) to a low excitation mode of the nucleus.
To carry out a shell model type calculation with such a deformed basis, 
the broken
rotational symmetry (and the particle number conservation if necessary)
has to be restored. Projection generates a new basis in the
laboratory frame in which the Hamiltonian is diagonalized. 
Thus, the final diagonalization is carried out in a space with a very 
small size (usually in a dimension smaller than 100), 
which 
has 
a well-defined  
microscopic structure and 
allows 
a physical interpretation in terms of rotational bands and the
interaction between them. 

Quasiparticles defined in the deformed Nilsson + BCS calculations are the
starting point of the PSM. 
The set of multi-qp states for our shell model configuration
space is
\begin{eqnarray}
\ack\Phi_\kappa\rangle=\{ 
\ a^\dagger_{\nu}\ack 0\rangle,
\ a^\dagger_{\nu}a^\dagger_{\pi_1} a^\dagger_{\pi_2}\ack 0\rangle \},
\label{confn}
\end{eqnarray}
for odd-neutron nuclei, and
\begin{eqnarray}
\ack\Phi_\kappa\rangle=\{
\ a^\dagger_{\pi}\ack 0\rangle,
\ a^\dagger_{\pi}a^\dagger_{\nu_1}a^\dagger_{\nu_2} \ack 0\rangle \},
\label{confp}
\end{eqnarray}
for odd-proton nuclei,
where $a^\dagger$'s are the quasiparticle (qp) 
creation operators, $\nu$'s ($\pi$'s)
denote the neutron (proton) Nilsson quantum numbers which run over
low-lying orbitals and $\ack 0 \rangle$ is the
Nilsson+BCS vacuum (0-qp state).
In Eqs. (\ref{confn}) and (\ref{confp}), the low-lying 3-qp states 
selected for the many-body basis are those consisting of 1-qp plus a pair
of qp's from nucleons of another kind. 
This selection is based on physical considerations.
In general, 3-qp states made by three nucleons of the same kind are also 
allowed, but such states usually lie higher in energy. 

The PSM employs the Hamiltonian \cite{Hara95}
\begin{equation}
\hat H = \hat H_0 - {1 \over 2} \chi \sum_\mu \hat Q^\dagger_\mu
\hat Q^{}_\mu - G_M \hat P^\dagger \hat P - G_Q \sum_\mu \hat
P^\dagger_\mu\hat P^{}_\mu.
\label{hamham}
\end{equation}
It contains the spherical single-particle term $\hat H_0$ 
which includes a proper spin-orbit force taken from Ref. \cite{BenRag}, 
the Q-Q interaction (the second term), and the monopole and
quadrupole pairing interactions (the last two terms). 
They represent the most important correlations in nuclei \cite{Zuker96}.
The Q-Q interaction strength $\chi$ is adjusted by the self-consistent
relation between the quadrupole deformation $\varepsilon_2$ and
the one resulting from the HFB procedure 
\cite{Hara95,Victor98}. The monopole
pairing strength $G_M$ is taken to be $G_M=\left[g_1\mp
g_2(N-Z)/A\right]/A$, where $g_1=22.5$ and $g_2=18.0$, and $-(+)$ is for
protons (neutrons). These are the appropriate values 
when three major shells 
($N=1,2,3$) 
are
included in the single-particle space \cite{Hara99}.
The quadrupole pairing strength $G_Q$ is assumed to be proportional to
$G_M$, with a proportionality constant taken in the range of 0.16 --
0.20.

For each spin $I$ the set of eigenvalue equations in the PSM are 
\cite{Hara95}
\begin{equation}
\sum_{\kappa'}\left\{H^I_{\kappa\kappa'}- 
E^{I \alpha} N^I_{\kappa\kappa'}\right\}
F^{I \alpha}_{\kappa'}=0, 
\label{psmeq}
\end{equation}
with $\alpha$ denoting states having a same spin $I$.
The Hamiltonian matrix elements $H^I_{\kappa\kappa'}$ and the norm
matrix elements $N^I_{\kappa\kappa'}$ are defined as
\begin{equation}
H^I_{\kappa\kappa'}=\langle\Phi_\kappa\ack\hat H\hat P^I_{KK'}\ack
\Phi_{\kappa'}\rangle,~~N^I_{\kappa\kappa'}=\langle\Phi_\kappa\ack\hat
P^I_{KK'}\ack\Phi_{\kappa'}\rangle,
\label{elem}
\end{equation}
where $\hat P^I_{MK}$ is the angular momentum projection operator.

The band energy is defined as the expectation value of the Hamiltonian
with respect to a ``rotational band $\kappa$''
$H^I_{\kappa\kappa}/N^I_{\kappa\kappa}$, 
which are the diagonal elements in Eq. (\ref{elem}).
A band diagram is a plot of the band energies as functions of spin $I$. 
It provide a useful tool for interpreting the results {\cite{Hara95}}.

\section{Band Staggering and the Signature Rule} 

Signature is a quantum number specifically appearing in a deformed
intrinsic system \cite{BM75}. 
It corresponds to a ``deformation invariance" of a system with quadrupole
deformation, thus it is not a direct concept
in the conventional spherical shell model.
Under a rotation of 180$^o$ around an axis perpendicular to the
symmetry axis, signature takes two eigenvalues.  
For an odd-$A$ nucleus, it is customary to assign 
\begin{equation}
\alpha_I = {1\over 2} (-1)^{I-1/2}
\end{equation}
as the signature quantum number for a state of spin $I$.
Thus, a rotational band with a sequence of levels differing in spin by
1 is now divided into two branches, each consisting of levels
differing in spin by 2 and classified by the signature quantum number
$\alpha_I = \pm {1\over 2}$, respectively. 
Experimentally, one often observes an energy staggering in rotational bands
and refers to this as signature splitting \cite{Hara92}. 

The origin of energy staggering in a rotational band can be
well explained by the particle-rotor model \cite{Hara95,Hara92}.
In a band associated with the last (decoupled) particle
having angular momentum $j$, 
the signature rule for a state with
total angular momentum $I$ 
can be expressed as 
\begin{equation}
I-j = \pm \left\{ 
\begin{array}{l} 
\mbox{even~~~~favored band}\\
\mbox{odd~~~~~unfavored band.} 
\end{array} \right.
\label{rule}
\end{equation}
Here, the favored band means the branch that is pushed down in energy, 
while the unfavored band the one that is pushed up.

However, as an observable phenomenon, splitting of one band into two
branches should also manifest itself in any nuclear many-body theory.
In fact, the $pf$ shell model diagonalization for $A=47$ and 49 nuclei
\cite{Pin97} has correctly reproduced the energy staggerings in the
rotational bands. 
 
In a theory based on angular momentum projection, one cannot explicitly
separate the angular momentum of the rotating body and that of the
(quasi)particles as in the particle-rotor model. 
Therefore, there is no ``decoupling parameter" \cite{BM75}
which splits a band into two by shifting the neighboring spin states
up or down.  
However, it was shown \cite{Hara95} that symmetrized wave functions 
exist in the PSM and the phenomenon of signature splitting 
shows up clearly when
an
intrinsic state is projected onto
a state of
good angular momentum. 

The PSM description of the signature splitting has some particular
features which are worth to be mentioned \cite{Sun94}.
The first one is the presence of signature dependence in bands with
the intrinsic quantum number $|K|>{1\over 2}$. It contrasts with
the particle-rotor model results, in which signature
splitting is confined only in the $K={1\over 2}$ case 
at first order in perturbation theory.
Nevertheless, for bands with $|K|>{1\over 2}$, 
splitting amplitude diminishes rapidly  as $|K|$ increases. 
The second PSM prediction is the increase in the splitting
amplitude with increasing $j$, and that bands
based on neighboring $j$-orbitals have opposite splitting phases.  
The third finding is related to the previous one: within a
band, the splitting amplitude is small when spin is low, but 
it increases with increasing spins.  
All these features have been extensively studied by 
Sun {\it et al.} \cite{Sun94}. 
As we shall see in the following discussion, these features are important
in explaining the experimental results.

\section{Structure of the Odd-Mass Nuclei}

In the present work, we study the odd-mass $pf$ shell
nuclei 
by applying the PSM in two sets of mirror nuclei:
$^{47}$V - $^{47}$Cr and $^{49}$Cr - $^{49}$Mn, which are odd-mass nuclei 
relative to the even-even $^{48}$Cr with either   
adding or removing one particle. In addition, we study
another two odd-mass nuclei $^{49}$V and $^{51}$Mn.

Given that these odd-mass nuclei are reasonably good rotors 
at least at low spins
\cite{Pin97},
we fix the same 
quadrupole deformation $\epsilon = 0.25$ in the deformed basis 
for all nuclei studied in this 
paper, which is the same value used in $^{48}$Cr \cite{Hara99}. 
We use
also the same monopole and quadrupole pairing strength constants
employed in the previous PSM calculations \cite{Hara99}.

In Fig 1, we present the energy spectra of these six nuclei. 
Comparison is made between the experimental data \cite{ensdf} shown at
the left hand side,  and the PSM results at the right hand side.
Fig. 2 contains the energy difference $E(I)-E(I-1)$ (in MeV) 
versus spin $I$ plots. The solid lines
represent the experimental data and the dashed lines the PSM results.
Fig. 3 shows the band diagrams $H^I_{\kappa\kappa}/N^I_{\kappa\kappa}$
versus spin $I$, with solid lines representing 1-qp bands, 
dashed lines 3-qp bands, 
and the diamonds the yrast band obtained from the PSM diagonalization
(experimental results are not shown in this figure).
In Fig. 3, although more low-lying bands have been included in the
calculation, we display only the most important bands to
illustrate the physics. Theoretical energies of the yrast bands in
Fig. 3 are those used in Figs. 1 and 2 to compare with data.

\subsection{$^{47}$V - $^{47}$Cr}

The pair of mirror nuclei $^{47}$V-$^{47}$Cr has been studied
within the {\em pf} shell model \cite{Pin97}, 
motivating new experimental work in 
this mass region \cite{Cam98}. 
These two sets of data look very similar, characterized by a nearly 
degenerate energy triplet near the ground state and several doublets 
above it, indicating strong energy staggerings in the yrast bands. 

Fig. 1a and 1b present the energy levels of $^{47}$V and $^{47}$Cr
respectively. 
The correct ground state spin and parity 
${\frac{3}{2}}^-$ are found for both nuclei. 
The almost degenerate ground-state triplet $[{\frac{3}{2}}^-,
{\frac{5}{2}}^-, {\frac{7}{2}}^-]$ is clearly reproduced. However, 
for $^{47}$Cr, the theory predicts an inversion between the
first ${\frac{5}{2}}^-$ and ${\frac{7}{2}}^-$ states which is not found in the
experiment. The PSM description of the spectra is quite good, but the
predicted moments of inertia are smaller than the experimental ones. It
can be clearly seen that for energies higher than 6 MeV, the predicted
states are displaced at higher energies compared to the experiment. 
Given that the Nilsson single particle energies do not show crossings in
this region around $\epsilon = 0.25$, we do not expect that 
increasing the mean
field deformation will modify the moment of inertia very much. 
Nevertheless, 
it is remarkable that the conceptually and numerically simple PSM
results are competitive with those found in the full {\em pf} shell model
diagonalizations \cite{Pin97}, which reproduced 
the overall energy scale very well, 
but predicted a couple of doublets inverted.

As can be seen in Fig. 2a,
the observed energy staggering is well reproduced in $^{47}$V. 
Similar staggering is predicted for $^{47}$Cr in Fig. 2b, 
for which 
the existing data
at low spins
are not sufficient to make a comparison. 
In both cases, the first two energy differences are exaggerated by the model. 

We now discuss the band diagram in Fig. 3 to understand the physics. 
In Fig. 3a, we see that the low spin states of the yrast band in $^{47}$V are
dominated by the proton $K=\frac{3}{2}$ state in $1f_{7/2}$ orbit.  
For $^{47}$Cr, the yrast band at low spins is mainly the band corresponding 
to the neutron $K=\frac{3}{2}$ state in $1f_{7/2}$ orbit.
Bands lying higher
in energy
have negligibly small influence in the mixing to the
$K=\frac{3}{2}$ one, as we can see that the filled diamonds 
take the value of the $K=\frac{3}{2}$ band almost exactly. 
The staggering in both yrast bands is such that the spin sequence of 
$I=\frac{3}{2}, \frac{7}{2}, ...$ is pushed down in energy, 
which follows exactly the rule of (\ref{rule})
with $j = 7/2$.  
Note that the physics is almost entirely described by projection onto the 
intrinsic 
$K=\frac{3}{2}$ state. The correct staggering phase and amplitude along
the bands are obtained by quantum mechanical treatment, 
in contrast to
the particle-rotor model results that no energy staggering should exist  
for the $K>{1\over 2}$ cases,
when the first order perturbation is introduced.

We thus understand why the experimentally observed energy doublets and
triplets can happen in these nuclei. 
Near the band head, the small splitting effect pushes the $I=\frac{7}{2}$ state 
close to the $I=\frac{3}{2}$ and $I=\frac{5}{2}$ states, forming 
a nearly degenerate energy triplet. 
Above them, states belonging to the $I=\frac{3}{2}$ branch are pushed down
to be close to the signature partners in the $I=\frac{5}{2}$ branch, 
so that the energy doublets are formed. In the $^{47}$V data, 
the energy difference between the states of $I=\frac{9}{2}$ 
and $I=\frac{11}{2}$,
and of $I=\frac{13}{2}$ and $I=\frac{15}{2}$, is close to zero, 
indicating an almost perfect degeneration. 
We notice that in the measured data of $^{47}$V and $^{47}$Cr, 
for several predicted degenerate doublets, only one state of the 
two signature partners 
has been observed.  

Data of the $^{47}$V and $^{47}$Cr nuclei seem to suggest 
that there is no band crossing at low spin states. 
From Fig. 3a and 3b we see that,   
in both nuclei, there is a band crossing near $I=\frac{19}{2}$
between the 1-qp $K=\frac{3}{2}$ band (lower solid line) and the 3-qp bands
(dashed lines). 
However, 
the band crossing in $^{47}$Cr occurs in higher energy without involving the 
lowest lying $K=\frac{3}{2} [1f_{7/2}]$ band.  
In $^{47}$V, the 3-qp bands cross the $K=\frac{3}{2} [1f_{7/2}]$ band
with only a small crossing angle so that the yrast band changes
smoothly its structure from the 1-qp to the 3-qp states.  
Therefore, in both cases, no sizable effect from the band
crossing can be observed in the energy spectra. 
This is in contrast to the situation in $^{49}$Cr and $^{49}$Mn nuclei,
as we shall discuss below.

\subsection{$^{49}$Cr - $^{49}$Mn}

The second pair of mirror nuclei studied in this paper is $^{49}$Cr-$^{49}$Mn.
For these two nuclei, the ground state spin and parity are $\frac{5}{2}^-$.
The separation of the energy doublets is generally larger, 
and the degeneracy is not as good as in $^{47}$V - $^{47}$Cr.  
In other words, the staggering amplitude is smaller at the low
spin states. 
Another notable differences from $^{47}$V - $^{47}$Cr are the
irregularities clearly seen in the spectra around the state
$I=\frac{17}{2}$.
The structure of this pair of nuclei have also been extensively 
discussed 
within the {\em pf} shell model \cite{Pin97}. 

The energy spectra shown in Fig. 1c and 1d agree well with the
experimental data. In particular, 
the doublet $[{\frac {25} {2}}^-, {\frac {27} {2}}^-]$
in $^{49}$Cr is predicted at energies above 10 MeV, well reproducing the 
experimental energies reported in the compilation of experimental data 
\cite{ensdf}. 
%It cast some doubts about the reinterpretation of these levels proposed
%in \cite{Pin97}. 
The irregular staggering in these two nuclei 
can be seen in Fig. 2c and 2d. The PSM reproduces fairly well the changes
in energy differences, which is more evident for $^{49}$Cr where there
are more data available. 

From Fig. 3c and 3d, we can see that 
in both nuclei, the yrast band is dominated by a
$K =\frac{5}{2}$ 1-qp band up to $I=\frac{15}{2}$. 
Since these are higher $K$-bands, the staggering is weaker than what we
have seen for the $K =\frac{3}{2}$ bands in Fig. 3a and 3b. 
This explains why the level separation of the energy doublets 
is generally larger in these two nuclei, as seen in Fig. 1. 

Around $I=\frac{17}{2}$, 
a 3-qp band crosses sharply the $K =\frac{5}{2}$ 1-qp band in both cases. 
This sharp crossing disturbs suddenly the regular band and changes 
components in the wave function. 
After the band crossing, the yrast bands are mainly 3-qp states. 
The 3-qp states in $^{49}$Cr and $^{49}$Mn have also mirror configurations. 
In $^{49}$Cr, it consists of the neutron $\frac{5}{2} [1f_{7/2}]$ plus 
a proton pair $K = 1 \{\frac{3}{2}[1f_{7/2}],\frac{5}{2}[1f_{7/2}]\}$, 
whereas in $^{49}$Mn, the structure is the proton 
$\frac{5}{2} [1f_{7/2}]$ plus
a neutron pair $K = 1 \{\frac{3}{2}[f_{7/2}],\frac{5}{2}[1f_{7/2}]\}$. 
Neglecting the effect of isospin 
symmetry breaking,
we thus expect 
the spectra of these two nuclei 
to be similar. 

Influence of the band crossing around $I=\frac{17}{2}$ in this pair of nuclei 
can be seen in Fig. 2c and 2d. 
The staggering changes in amplitude as function of spin are 
correctly
reproduced, although there are visible deviations.

\subsection{$^{51}$Mn and $^{49}$V}

As an isotope of $^{49}$Mn, we expect a similar behavior of the 
lowest proton 1-qp state 
in $^{51}$Mn.  
One may also expect to see differences around the band crossing region,
where two additional neutrons are involved in the 3-qp states. 
Due to shift of the neutron Fermi levels in the two nuclei
$^{49}$Mn and $^{51}$Mn, behavior
of the 3-qp bands can be different, which can leave the energy spectra
around the band crossing regions different. 

The energy spectra of $^{51}$Mn and $^{49}$V are shown in Fig. 1e and 1f. 
As in the $A=47$ case, the ordering of the states 
as well as the ground state spin are well reproduced.
As shown in Fig. 2e, 
the PSM reproduces the energy differences  
with correct phases and amplitudes for the 
most measured states in $^{51}$Mn, but 
fails to reproduce the sudden reduction 
of the $E(I)-E(I-1)$ value at spin $I=\frac{19}{2}$.
The reason for this discrepancy can be seen in Fig. 3e, where the position 
of the 3-qp
band is too high in energy, and thus crosses the 1-qp $K=\frac{5}{2}$ 
band gently at 
a too late spin state. 

In Fig. 2f, the staggering in $^{49}$V is successfully described
by the PSM
up to $I=\frac{19}{2}$. For larger spin states 
the PSM underestimates the changes in energy differences. 
This discrepancy, as can be seen in Fig. 3f, has its origin in the fact that 
the 3-qp state consisting of the proton $\frac{3}{2} [1f_{7/2}]$ plus
the neutron pair $K = 3 \{\frac{1}{2}[2p_{3/2}],\frac{5}{2}[1f_{7/2}]\}$ 
dives into the yrast region after the band crossing. This is a high-$K$
band, thus has a very weak staggering effect, acoording to our early
discussions. 

There is an intersting observation that is worth to be pointed out.
As an isotope of $^{47}$V, we expect that the 1-qp proton state 
$\frac{3}{2} [1f_{7/2}]$ dominates the low spin states in the yrast band
of $^{49}$V also, and that similar triplet states should be observed around
the ground state. Indeed, such a nearly degenerate triplet has been
seen experimentally as shown in Fig. 1f, but the order of these states
are reversed as  $[{\frac{7}{2}}^-,
{\frac{5}{2}}^-, {\frac{3}{2}}^-]$.  
It is seemingly anomalous that this nucleus has the ground state spin
$I=\frac{7}{2}$, but the corresponding intrinsic state has $K=\frac{3}{2}$.   
The physical reason for this anomaly can be understood as the interaction
of the 1-qp proton $\frac{3}{2} [1f_{7/2}]$ band with the 1-qp proton
$\frac{5}{2} [1f_{7/2}]$ band which lies just little above the former. 
As the consequece of the interaction, 
the yrast states of $I=\frac{5}{2}$ and $\frac{7}{2}$
are pushed down in energy, and become lower than  
the $I=\frac{3}{2}$ state which should
otherwise be the ground state as in the $^{47}$V case. 
This feature has been described correctly by the PSM.

\section{Summary}

The present results confirm that the PSM is a practical tool in studying 
rotational bands, which, having been successful in describing deformed and
triaxial rare earth nuclei, is also able to provide a quantitative and
simple description of light nuclei. 
In a previous work, the backbending of
$^{48}$Cr was described using the PSM as a band crossing phenomena 
found in heavier nuclei. The same ideas have proved fruitful
here to describe the 
irregularities 
in rotational bands in odd-mass nuclei.
In addition, we have discussed the energy staggering in these bands, 
in terms of the signature splitting. 

While the description of the energy spectra of these nuclei is in general
very good, there is a general tendency in the model 
that the moment of inertia is underestimated. 
Having checked that changes in the mean field
deformation do not induce changes in the moment of inertia in these nuclei
when the PSM is employed, we could speculate that a monopole pairing
strength is excessively strong and could be responsible for this effect.
Modifying this interaction would have an important effect in the spectra 
with only minor changes in the yrast wave functions \cite{Zuker95}. In any
case, we preferred to leave the PSM parameters unchanged, {\em i.e.} use
the same employed in the calculation of $^{48}$Cr,
to exhibit the consistence of the description for even- and odd-mass
nuclei in this mass region. 

V.V. is a fellow of Conacyt (Mexico). 
Partial 
financial support from Conacyt is acknowledged. 
  
\newpage
\baselineskip = 16pt
\bibliographystyle{unsrt}

\begin{thebibliography} {99}
\bibitem{Cam94} J.A. Cameron et al., Phys. Rev. {\bf C49} (1994) 1347.
% exp. 47Ti,V,Cr, 48V,Cr 
\bibitem{Cam96} J.A. Cameron et al. Phys. Lett. {\bf B235} (1990) 239;
Phys. Lett. {\bf B319} (1993) 58; Phys. Lett. {\bf B387} (1996) 266.
\bibitem{Cam98} J.A. Cameron et al., Phys. Rev. {\bf C58} (1998) 808.
% exp. 46Ti, 50Cr, 47V, 49Cr 
\bibitem{Lenzi97} S.M. Lenzi et al., Phys. Rev. {\bf{C56}} (1997) 1313.
% exp 50Cr, 2nd backbending
\bibitem{Bra98} F. Brandolini et al., Nucl. Phys. {\bf A642} (1998) 387.
% exp. 48,50Cr
\bibitem{Cau94} E. Caurier, A.P. Zuker, A. Poves and  G. Mart\'\i nez-Pinedo,
Phys. Rev. {\bf C50} (1994) 225.
% SM  A=48
\bibitem{Cau95} E. Caurier, J.L. Egido, G. Mart\'\i nez-Pinedo, A. Poves, 
 J. Retamosa, L.M. Robledo, and A.P. Zuker, Phys. Rev. Lett. {\bf 75} (1995)
2466. % SM y HFB 48Cr
\bibitem{Pin96} G. Mart\'\i nez-Pinedo, A. Poves, L.M. Robledo, E. 
Caurier, F. Nowacki, J. Retamosa and A. Zuker, Phys Rev {\bf C54} (1996)
R2150.
% Cr50 backbending
\bibitem{Pin97} G. Mart\'\i nez-Pinedo, A.P. Zuker, A. Poves and 
E. Caurier, Phys. Rev. {\bf C55} (1997) 187.
% SM  A=47, 49
\bibitem{Rin80} P. Ring and P. Schuck, {\it The Nuclear Many Body Problem} 
(Springer Verlag, Berlin, 1980).
\bibitem{Tan98} T. Tanaka, K. Iwasawa and F. Sakata, Phys. Rev.  
{\bf C58} (1998) 2765. % CHFB, NO s.p. level crossing
\bibitem{Hara99} K. Hara, Y. Sun and T. Mizusaki, Phys. Rev.
Lett. {\bf 83} (1999) 1922. % PSM 48Cr
\bibitem{Hara95} K. Hara and Y. Sun, Int. J. Mod. Phys. {\bf{E 4}} 
(1995) 637. % PSM Review
\bibitem{Shei99} J.A. Sheikh and K. Hara, Phys. Rev. Lett. {\bf 82} (1999)
3968. 
\bibitem{BenRag} T. Bengtsson and I. Ragnarsson, Nucl. Phys. {\bf A436},
(1985) 14.
\bibitem{Zuker96} M. Dufour and A.P. Zuker, Phys. Rev. {\bf C54} (1996)
1641.
\bibitem{Victor98} V. Vel\'azquez, J. Hirsch and Y. Sun,
Nucl. Phys. {\bf A643} (1998) 39.
\bibitem{BM75} A. Bohr and B.R. Mottelson, {\it Nuclear Structure} 
 (Benjamin, New York, 1975).
\bibitem{Hara92} K. Hara and Y. Sun, Nucl. Phys. {\bf A537} (1992) 77.
\bibitem{Sun94} Y. Sun, D.H. Feng and S.X. Wen, Phys. Rev. {\bf C50} (1994) 
2351.  
\bibitem{ensdf} ENSDF, http://ie.lbl.gov/ensdf/welcome.htm . 
\bibitem{Zuker95} A.P. Zuker. J. Retamosa, A. Poves and E. Caurier, Phys.
Rev. {\bf C52 } (1995) R1742.

\end{thebibliography}

%\newpage
\begin{figure}
\caption{
Energy levels of $^{47}$V, $^{47}$Cr, $^{49}$Cr, $^{49}$Mn,
$^{51}$Mn and $^{49}$V are shown in inserts a-f  respectively.
Experimental levels are presented in the left hand side and the PSM
results at right hand side.
}
\label{figure.1}
\end{figure}

\begin{figure}
\caption{
$E(I)-E(I-1)$ vs. $I$ curves with the same order of Fig. 1.
The experimental values are plotted in solid lines with diamonds, the PSM
calculations in dashed lines with crosses.
}
\label{figure.2}
\end{figure}

\begin{figure}
\caption{
Band diagrams  $E$ vs. $I$ with the same order of Fig.
1. The yrast band is represented with diamonds, the 1qp bands with solid
lines and the  3qp bands with dashed lines.
}
\label{figure.3}
\end{figure}

\end{document}